# On Access Control in Cabin-Based Transport Systems

Pasquale Grippa, Udo Schilcher and Christian Bettstetter

*Abstract*—We analyze a boarding solution for a transport system in which the number of passengers allowed to enter a transport cabin is automatically controlled. Expressions characterizing the stochastic properties of the passenger queue length, waiting time, and cabin capacity are derived using queuing theory for a transport line with deterministic arrivals of cabins and Poisson arrivals of passengers. Expected cabin capacity and stability threshold for each station are derived for a general passenger arrival distribution. Results show that a significant reduction of the waiting time at a given station is only possible at the cost of making the stability of one of the preceding stations worse than that of the given station. Experimental studies with real passenger arrivals are needed to draw firm conclusions.

## I. Introduction

A novel boarding solution for cabin-based transport systems (e.g., ski lifts, cable cars, subways) is being discussed in the industry and has already been implemented in the Austrian skiing resort Bad Gastein [1]: In order to avoid long queues at succeeding boarding stations, a display in the boarding area tells the guests how many of them are allowed to enter the next cabin. This form of access control guarantees spare seats for passengers waiting at the middle station to go to the top station. The overall objective is to install fair access conditions at all stations, which would automatically improve waiting time and comfort of passengers. The operator expects some system intelligence to compute the number of passengers to enter at each station and adapt this number in real time according to the varying passenger load. Compared to extensions or modifications of tracks and cabins, access control would be an inexpensive solution to optimize systems.

This article assesses as to whether such access control can actually improve the service for a system with a simple linear topology of stations. Queuing theory is applied to analyze the impact on waiting time and system stability. The arrival of passengers at each station is stochastic: We assume a Poisson arrival process and a system in stationary conditions (fixed arrival rate). The results can be summarized as follows:

- It is impossible to significantly reduce the waiting time of a given station $m$ by controlling the access of passengers at previous stations without making the stability of one of the previous stations worse than the stability of station $m$. This is valid for a general arrival distribution.
- It is possible to achieve a moderate improvement at station $m$ to an extent that is relevant in real operation but at the cost of greatly decreasing the stability of a preceding station.
- A main cause of improvement is the reduction of the variance in the cabin capacity. Therefore, access control policies should focus on reducing the variance.

The article is structured as follows: Section II addresses related work in the domain of queuing theory. Section III specifies the transport system, including modeling assumptions and quantities of interest, such as queue length, waiting time, and stability threshold. Section IV defines the research question and explains key results. In Section V, we derive the spectral representation of the waiting time and queue length at a station, the probability distributions of quantities describing the interaction between stations, and the stability threshold. Section VI applies this theoretical framework to a gondola lift in a ski resort. Section VII concludes with a suggestion on how to design good access control policies.

## II. Related Work

It follows an overview of results on queuing systems with Poisson arrivals where the queue is served in batches with a given maximum size and a random independently distributed time between consecutive services. For systems that only have a single queue, expressions for the first two moments of the queue length ([2], [3]) and the waiting time distribution ([4], [5]) are known. Furthermore, a time dependent solution of this model is available [6]. Since these results are complex, approximations for the mean queue length for numerical evaluation have been derived [7]. These approximations are simple in comparison to the exact solution yet achieve accurate results. All this work assumes that every time the queue is long enough, the server can serve a batch of fixed size, independent of other circumstances. This is, however, unrealistic for certain transport system, e.g., elevators, where part of the capacity may be occupied when arriving at a floor. Hence, the basic queuing model has been extended accordingly [8] and the equilibrium distribution of the queue length has been derived for this extended model.

A further generalization is to consider that the time between services follows a general distribution. For this model the distribution of the number of passengers waiting can be derived [9]. For general arrivals and exponentially distributed times between services the waiting time distribution has been

P. Grippa and C. Bettstetter are with the Institute of Networked and Embedded Systems, Alpen-Adria-Universität Klagenfurt, Austria. U. Schilcher is with Lakeside Labs GmbH, Klagenfurt, Austria. Email: pasquale.grippa@aau.at. This work was supported by Universität Klagenfurt and Lakeside Labs GmbH, Klagenfurt, Austria and funding from the European Regional Development Fund and the Carinthian Economic Promotion Fund (KWF) under grant KWF 20214 a. 3520/ 27678/39824.

derived [10], [11]. Furthermore, for bulk-arrival bulk-service queues, the moments of the queue length distribution [12] and the equilibrium waiting-time distribution [13] are known. When the queue length is limited, results similar to the above are harder to obtain. Some results are available, e.g., on the waiting time [14] and the queue length [15]. This type of model is, however, not developed far enough to be applied to the scenario considered in the paper at hand.

## III. SYSTEM MODEL AND DEFINITIONS

### A. General System Description

The system is composed as shown in Fig. 1. Stations are located along a closed path (station line). Cabins are uniformly spaced, move along the line, and stop at stations at constant time intervals $\beta$; each cabin has $\gamma$ seats (cabin size). At each station, the passengers in the cabin can leave and, afterward, those waiting at the station can enter. Each station is divided into three areas: waiting area, boarding buffer, and platform. Before the cabin arrives, the boarding buffer is filled with the passengers to be served (batch). All passengers waiting in the buffer board during the current service, so that the buffer gets empty for the next service. At the cabin arrival, the gate between waiting area and boarding buffer is closed, so that new passengers cannot join the ongoing service. If the number of waiting passengers is smaller than the available free seats (cabin capacity without access control), the cabin does not wait for passengers to arrive but leaves the station (no waiting for the batch formation). The service of a batch starts with the arrival of a cabin and ends with the arrival of the next cabin (a new batch can be served). Thus, the service time is constant and equal to the interarrival time of cabins $\beta$.

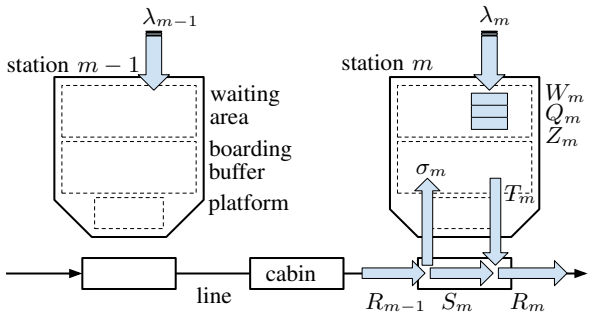

Figure 1: Model of the transport system

### B. Mathematical Notation

The following notation is used to describe the system behavior at a given station $m$ in a formal way:
- Parameters of the system are Greek letters (e.g., $\lambda_m$);
- random variables and stochastic processes are uppercase Latin letters (e.g., $X_m$ and $\{X_{m,n}, n \geq 1\}$), with $X_m = \lim_{n \to \infty} X_{m,n}$;
- probability distribution functions (continuous variables) and probability mass functions (discrete variables) are lowercase Latin letters (e.g., $x(\cdot)$);
- Laplace–Stieltjes (LS) transforms (continuous variables) or probability generating functions (discrete variables) are uppercase Latin letters with asterisk (e.g., $X^*(\cdot)$).

### C. Queue Length and Waiting Time

Passengers arrive at station $m$ according to a Poisson arrival process with rate $\lambda_m = \lambda \nu_m$, where $\lambda$ is the total arrival rate of the system and $\nu_m$ is the fraction of these passengers arriving at station $m$. Note that the Poisson assumption is not required for the derivation of the stability thresholds and expected cabin capacities. In stationary conditions, a passenger arriving at station $m$ finds a queue of $Q_m$ waiting passengers (queue length), including passengers in the batch, and waits $W_m$ time units (waiting time). These random variables depend on the number of passengers $C_m$ that can access the cabin at this station (cabin capacity). Each of the $R_{m-1}$ passengers riding from the previous station $(m-1)$ leaves the cabin at station $m$ with probability $\sigma_m$, and thus makes a seat free for the waiting passengers. The remaining $S_m$ passengers stay in the cabin because they want to travel further. In general, we can write that the capacity is $C_m = g_m(S_m)$, where $g_m(\cdot)$ is the access control policy at station $m$. An example policy is to limit the maximum number of accesses per cabin to $\eta_m \in [1, \gamma]$, i.e. $C_m = \min[\eta_m, \gamma - S_m]$. If no access control is applied, the capacity is equal to the number of free seats in the cabin, i.e. $C_m = \gamma - S_m$. A cabin arriving in station $m$ finds $Z_m$ passengers waiting in the queue (passengers left behind upon the departure of the previous batch). Of those, $T_m$ enter the cabin, so that $R_m = S_m + T_m$ ride to the next station $m + 1$. Note that the random variables $Q_m$ and $Z_m$ are different because they are observations of the same continuous-time stochastic process at different time instances.

Our goal is to obtain stochastic descriptions of $W_m$ and $Q_m$ under static access control. The probability mass function of the queue length can be computed from the probability generating function $Q_m^*(\cdot)$ by

$$q_m(k) = \frac{1}{k!} \left. \frac{\mathrm{d}^k Q_m^*(z)}{\mathrm{d}z^k} \right|_{z=0}. \quad (1)$$

Furthermore, we are interested in the moments of the waiting time, which can be computed from the LS transform $W_m^*(\cdot)$:

$$\mathrm{E}\left[W_m^k\right] = (-1)^k \left. \frac{\mathrm{d}^k W_m^*(s)}{\mathrm{d}s^k} \right|_{s=0}. \quad (2)$$

### D. Stability and Fairness

A station is stable if the expected waiting time is finite. This holds if the arrival rate $\lambda_m$ is smaller than the stability threshold $\lambda_m^* = \mathrm{E}\left[C_m\right]/\beta$. In terms of the load factor $\rho_m$, we require $\rho_m = \lambda_m \beta / \mathrm{E}\left[C_m\right] < 1$. A station is said to be in low load if $\lambda_m$ is much smaller than $\lambda_m^*$; in this case, there is almost always no queue and the expected waiting time is $\mathrm{E}\left[W_m\right] = \beta/2$. A station is said to be in high load if $\lambda_m$ is close to $\lambda_m^*$; in this case, $\mathrm{E}\left[W_m\right]$ changes dramatically for small variations of $\lambda$.

All stations on a given line share the capacities of the cabins. It is desirable to have a fair allocation of cabin capacities among stations, so that the expected waiting time $\mathrm{E}\left[W_m\right]$ is in the same order of magnitude for all stations. If we exclude the trivial case with all stations being in low load ($\mathrm{E}\left[W_m\right] \simeq \beta/2$), having similar expected waiting times is equivalent to have similar scaled stability thresholds $\lambda_m^*/\nu_m$.

## IV. PROBLEM STATEMENT AND KEY RESULTS

The capacity of cabins arriving at a particular station can be modified by controlling the access of passengers at preceding stations. A fundamental question in this context is: Is it possible to significantly decrease the expected waiting time of a station by controlling the access of passengers at the preceding stations that have a better stability threshold?

We show that this is impossible under the given modeling assumptions. To significantly improve the performance of a station $m$, we have to increase the scaled stability threshold $\lambda_m^*/\nu_m$ of that station. To do so, we have to reduce the scaled stability threshold $\lambda_k^*/\nu_k$ of at least one preceding station $k$ to a value smaller than $\lambda_m^*/\nu_m$ (Section V-C, Fig. 2 and Fig. 3). This makes the performance of $k$ similar to (capacity variance plays a role) or worse than the original performance of $m$. If $\lambda_k^*/\nu_k > \lambda_m^*/\nu_m$, a large reduction of $\lambda_k^*/\nu_k$ leads to a modest relative gain in the waiting time of $m$, which might be interesting in practice though (Section VI and Fig. 4). Since $\lambda_m^*/\nu_m$ does not change for variations of $\lambda_k^*/\nu_k$ as long as $\lambda_k^*/\nu_k > \lambda_m^*/\nu_m$ (see Fig. 2), the relative gain cannot be due to an increased capacity *mean* ($\mathrm{E}\left[C_m\right] = \lambda_m^*\beta$). It is a reduced capacity *variance* that brings the relative gain (Section VI and Fig. 5).

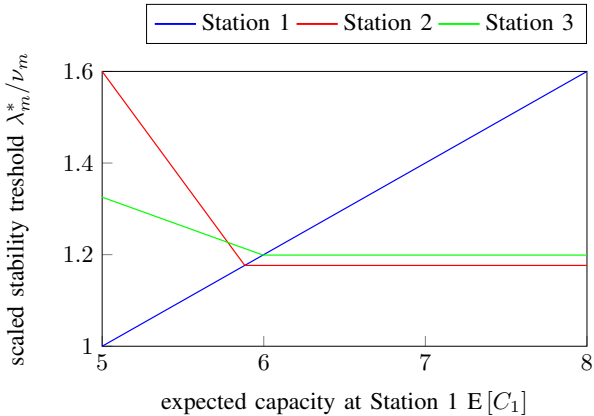

Figure 2: Stability thresholds at all stations as a function of the capacity at Station 1. Parameters: $\beta = 10$, $\gamma = 8$, $R_0 = 0$, $[\nu_1, \nu_2, \nu_3, \nu_4] = [0.5, 0.2, 0.3, 0]$, $[\sigma_1, \sigma_2, \sigma_3, \sigma_4] = [0, 0.04, 0.46, 1]$.

Fig. 2 illustrates this general result for a ski resort in Bad Gastein (Austria) with four stations in a line. The parameters characterizing the main line of the ski resort from real data are as follows: interarrival between cabins $\beta = 10\,\mathrm{s}$, cabin size $\gamma = 8$, fraction of passengers arriving at the stations $[\nu_1, \nu_2, \nu_3, \nu_4] = [0.5, 0.2, 0.3, 0]$, probability of leaving the cabin at the stations $[\sigma_1, \sigma_2, \sigma_3, \sigma_4] = [0, 0.04, 0.46, 1]$. The parameters $\nu_m$ are estimated using smart turnstiles and cameras, and include passengers that do not leave the system after one travel but stay in the system and come back for additional rides. The parameter $\sigma_3$ is high because Station 3 allows to change to another line. Passengers not leaving at this station travel to the final destination (Station 4). No passenger enters at Station 4. Since $\sigma_4 = 0$, cabins at Station 1 are always empty, i.e. $R_0 = 0$ and in turn $C_1 = \eta_1$. The scaled stability thresholds are described by (24), (25), and (26). Fig. 2 shows the scaled stability thresholds of all stations as a function of the expected capacity at Station 1. As the capacity decreases (starting from 8), $\lambda_1^*/\nu_1$ decreases, but $\lambda_2^*/\nu_2$ remains unchanged. $\lambda_2^*/\nu_2$ starts to increase only when $\lambda_1^*/\nu_1 < \lambda_2^*/\nu_2$. The same holds for $\lambda_3^*/\nu_3$. It is important to note that this result is independent of how the capacity is obtained, i.e., on the static access control policy applied.

## V. DERIVATION OF QUEUE LENGTH, WAITING TIME, AND STABILITY THRESHOLD

### A. Queue Length and Waiting Time at a Single Station

A station $m$ can be modeled as an M/D/1 queue with batch services (extensively studied in [11, Ch. III.2]). The probability generating function of the queue length is [11, Ch. III.2]

$$Q_m^*(z) = \frac{Z_m^*(z)}{A_m^*(z)} \frac{1 - A_m^*(z)}{(1 - 1/\xi)\lambda_m\beta} \;, \qquad (3)$$

and the LS transform of the waiting time is [11, Ch. III.2]

$$W_m^*(s) = \frac{1}{2\pi i} \oint_{\mathcal{C}} \frac{z}{(\xi - z)(1 - \xi)} \frac{1 - C_m^*(\xi)}{1 - B_m^*(s)C_m^*(\xi)}$$
$$\frac{Z_m^*(1/\xi)}{A_m^*(1/\xi)} \frac{B_m^*(s) - A_m^*(1/\xi)}{(1 - 1/\xi)\lambda_m\beta - s\beta} \,\mathrm{d}\xi \;. \quad (4)$$

The integration domine $\mathcal{C}$ is the unit circle centered at the origin of the complex plane; $A_m^*(\xi)$ is the probability generating function of the number of arrivals during a service, which is $\mathrm{e}^{(\xi-1)\lambda_m\beta}$ for Poisson arrivals; $B_m^*(s)$ is the LS transform of the service time, which is $\mathrm{e}^{-s\beta}$ for deterministic service; and $C_m^*(\cdot)$ and $Z_m^*(\cdot)$ are the probability generating functions of the variables introduced above in stationary conditions.

The capacity is represented by $C_m^*(z) = \sum_{j=0}^{\eta_m} c_m(j) z^j$, where $\eta_m \leq \gamma$ is the maximum number of free seats imposed by the access control. The values of the coefficients $c_m(\cdot)$ (probability mass function) depend on arrivals and access control policies at station $m$ and preceding stations interacting with station $m$ through the shared cabins (see Sec. III-C and Sec. V-B). Using the polar notation $\xi = \mathrm{e}^s$, the function $Z_m^*(\cdot)$ can be calculated by solving the limit [11, Ch. III.2]

$$Z_m^*(\mathrm{e}^{-s}) = \lim_{r \to 1^-} \left((1-r)\,\zeta_m(s,r)\right) \;\text{with} \qquad (5)$$

$$\zeta_m(s,r) = \sum_{n=2}^{\infty} r^n \,\mathrm{E}\left[\mathrm{e}^{-sZ_{m,n}}|Z_{m,1} = 0\right] \;, \qquad (6)$$

where the stochastic process $\{Z_{m,n}, n \geq 1\}$ represents the number of passengers left behind upon the departure of the $(n-1)$th batch service, i.e. the number of waiting passengers at the arrival time of the $n$th cabin at the station. For simplicity, since we evaluate the system in stationary conditions

$(n \to \infty)$, we drop the index $n$. Sum (6) can be expressed as the integral [11, Ch. III.2]

$$\zeta_m(s,r) = r^2 A_m^*(\mathrm{e}^{-s}) \tag{7}$$
$$\exp\left[-\frac{1}{2\pi i} \int_{\epsilon-i\infty}^{\epsilon+i\infty} \frac{s\,\mathrm{d}\xi}{\xi(s-\xi)} \log\left[1 - rA_m^*(\mathrm{e}^{-\xi})C_m^*(\mathrm{e}^{\xi})\right]\right]$$

with $\mathrm{Re}\,s > 0$, $|r| < 1$, and $0 < \epsilon < \mathrm{Re}\,s$, and $\epsilon$ sufficiently small to include all the poles of the logarithmic function. This integral can be solved by using Cauchy's integral formula and the residual theorem. Its value depends on the residuals of the integrand in $\xi = s$ and in the poles of the logarithm included in the semi-plane $\mathrm{Re}\,\xi > 0$. These poles are the solutions of the transcendental equation

$$1 - r\mathrm{e}^{(z-1)\lambda_m \beta} C_m^*(z^{-1}) = 0 \tag{8}$$

inside the unitary circle $|z| < 1$ with $z = \mathrm{e}^{-\xi}$. The solution of the integral is

$$\zeta_m(s,r) = \frac{r^2 A_m^*(\mathrm{e}^{-s})}{1 - rA_m^*(\mathrm{e}^{-s})C_m^*(\mathrm{e}^s)} \prod_{j=1}^{\eta_m} \frac{1 - \mathrm{e}^s \mu_j(1)}{1 - \mu_j(1)}, \tag{9}$$

where $\{\mu_j(r), j = 1, \ldots, \eta_m\}$ are the solutions of (8) in the unit circle. For $|r| < 1$, these roots are inside the unit circle. For $r \to 1^-$, the roots with $j \geq 2$ are inside the unit circle and $\mu_1(r) \to 1$. As $r \to 1^-$, $\zeta_m(s,r) \to \infty$, but the product $\zeta_m(s,r)(1-r)$ stays finite. To compute the limit (5) we have to find an analytic description of $\mu_1(r)$. For $r \to 1^-$, $\mu_1(r)$ is the $z$ that solves the equation

$$1 - rA_m^*(z)C_m^*(z^{-1}) = 0. \tag{10}$$

Expanding (10) into a power series around $z = 1$ and substituting $z$ with $\mu_1(r)$, we obtain

$$1 - r + r(\mathrm{E}\,[C_m] - \lambda_m\beta)(\mu_1(r) - 1) + o(\mu_1(r) - 1) = 0. \tag{11}$$

In the limit (5) ($r \to 1^-$), all the terms approaching zero faster than linearly ($o(\mu_1(r) - 1)$) can be neglected. Therefore,

$$1 - \mu_1(r) = \frac{1-r}{r(\mathrm{E}\,[C_m] - \lambda_m\beta)}. \tag{12}$$

Substituting $1 - \mu_1(r)$ into (9), taking the limit for $r \to 1^-$, and substituting $z = \mathrm{e}^{-s}$ yields the final result

$$Z_m^*(z) = \tag{13}$$
$$(\mathrm{E}\,[C_m] - \lambda_m\beta)\frac{(1-z^{-1})A_m^*(z)}{1 - A_m^*(z)C_m^*(z^{-1})} \prod_{j=2}^{\eta_m} \frac{1 - z^{-1}\mu_j(r)}{1 - \mu_j(r)}.$$

The numerical value of $\{\mu_j(r), j \geq 2\}$ can be found by expanding (8) into a series around zero, solving the relative polynomial equation, and selecting the values inside the unitary circle. Once $Z_m^*(z)$ is known, $W_m^*(s)$ is obtained by numerically solving the integral in (4).

*B. Interactions Among Stations*

The performance of a station $m$ depends on its capacity $C_m$, which is a function of arrivals and access control policies at station $m$ itself and at the preceding stations. The number of passengers in the cabins approaching station $m$ are distributed according to $r_{m-1}(\cdot)$. Each riding passenger leaves the cabin at station $m$ with probability $\sigma_m$ and gives a free seat for the passengers waiting at the station. Thus, the number of passengers staying in the cabin follows

$$s_m(k) = \sum_{j=k}^{\gamma} \binom{j}{k}(1-\sigma_m)^k \sigma_m^{j-k} r_{m-1}(j). \tag{14}$$

Special cases are $s_m(k) = r_{m-1}(k)$ for $\sigma_m = 0$, and $S_m = 0$ for $\sigma_m = 1$.

Access control limits the number of passengers in $m$ to have more free seats at station $m+1$. If we limit the maximum number of accesses per cabin, the capacity is $C_m = \min[\eta_m, \gamma - S_m]$ with $1 \leq \eta_m \leq \gamma$. The cumulative probability distribution is

$$\mathrm{P}\,[C_m \leq k] = \begin{cases} 1 & \text{for } k \geq \eta_m \\ 1 - \mathrm{P}\,[S_m \leq \gamma - k - 1] & \text{else.} \end{cases} \tag{15}$$

The actual number of passengers entering the cabin during the current service is $T_m = \min[Z_m, C_m]$. $Z_m$ is the number of passengers waiting at the station when the cabin arrives. Because of causality, this quantity cannot depend on the capacity of the current cabin but depends on the capacities of the cabins arrived in the past. Therefore the variables $Z_m$ and $C_m$ are independent. Given this independence, we have

$$\mathrm{P}\,[T_m \leq k] = 1 - \mathrm{P}\,[C_m > k]\mathrm{P}\,[Z_m > k]. \tag{16}$$

The number of passengers riding to the next station is $R_m = T_m + S_m$ with distribution

$$\mathrm{P}\,[R_m \leq k] = \mathrm{P}\,[T_m + S_m \leq k] \tag{17}$$
$$= \sum_{j=0}^{\gamma} \mathrm{P}\,[T_m \leq k-j | S_m = j]\mathrm{P}\,[S_m = j]$$
$$= \sum_{j=0}^{\gamma} (1 - \mathrm{P}\,[Z_m > k-j, C_m > k-j | S_m = j])\mathrm{P}\,[S_m = j]$$
$$= \begin{cases} 1 & \text{for } k = \gamma \\ \mathrm{P}\,[S_m \leq k - \eta_m] + \sum_{j=k-\eta_m+1}^{k} \mathrm{P}\,[Z_m \leq k-j]\mathrm{P}\,[S_m = j] \\ \quad \text{for } \eta_m \leq k < \gamma \\ \sum_{j=0}^{k} \mathrm{P}\,[Z_m \leq k-j]\mathrm{P}\,[S_m = j] & \text{for } k \leq \eta_m - 1. \end{cases}$$

An instability of station $m$ does not necessarily imply instability of all succeeding stations if $\eta_m < \gamma$ or $\sigma_i > 0$ for $i > m$. If $m$ is unstable, $T_m = C_m$ and

$$\mathrm{P}\,[R_m \leq k] = \begin{cases} 1 & \text{for } k = \gamma \\ \mathrm{P}\,[S_m \leq k - \eta_m] & \text{for } \eta_m \leq k < \gamma \\ 0 & \text{for } k \leq \eta_m - 1. \end{cases} \tag{18}$$

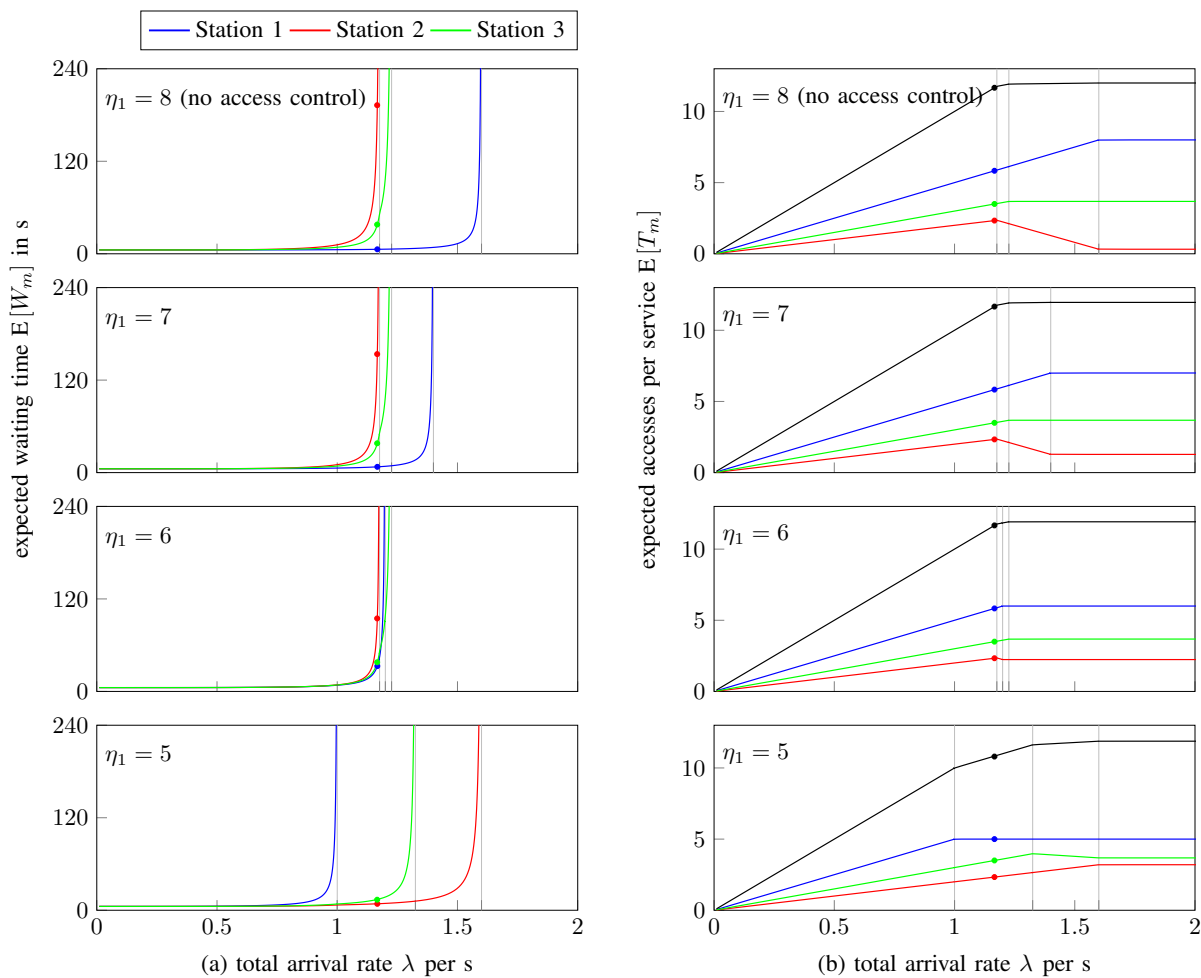

Figure 3: Effect on Station 2 and 3 of access control at Station 1. Parameters: $\beta = 10$, $\gamma = 8$, $R_0 = 0$, $[\nu_1, \nu_2, \nu_3, \nu_4] = [0.5, 0.2, 0.3, 0]$, $[\sigma_1, \sigma_2, \sigma_3, \sigma_4] = [0, 0.04, 0.46, 1]$. All results are obtained by mathematics; they show a very good fit with simulations (not shown).

## C. Stability Threshold

The performance of a station is limited by its stability threshold $\lambda_m^* = \mathrm{E}[C_m]/\beta$. The waiting time distribution is undefined for arrival rates above this threshold. The expected number of waiting passengers steadily increases over time; the system thus never reaches a stationary state. To significantly reduce the waiting time and improve its robustness to arrival rate variation, the stability threshold must be increased.

In stationary conditions (stable system), the expected number of passengers entering the station is equal to the expected number of passengers leaving it (entering the cabins), i.e., $\nu_m \lambda \beta$. A station stays stable if this quantity is smaller than the expected cabin capacity. Therefore, the expected number of passengers entering the cabin at station $m$ is

$$\mathrm{E}[T_m] = \min[\nu_m \lambda \beta, \mathrm{E}[C_m]]. \quad (19)$$

If no access control is applied at station $m$, we have

$$\mathrm{E}[C_m] = \gamma - \mathrm{E}[R_0] \prod_{i=1}^{m}(1-\sigma_i) - \sum_{j=1}^{m-1} \mathrm{E}[T_j] \prod_{i=j+1}^{m}(1-\sigma_i) \quad (20)$$

where the sum is zero for $m = 1$. Access control at a station $k < m$ reduces $\mathrm{E}[C_k]$, thus decreases $\mathrm{E}[T_k]$ and in turn increases $\mathrm{E}[C_m]$.

We refer to $\lambda_m^*/\nu_m$ as scaled stability threshold. We now show that it is impossible to increase the scaled stability threshold of a given station $m$ by controlling the access of passengers at the preceding stations with better stability without making the stability of at least one of the preceding stations worse than the stability of station $m$. Under the assumption that station $m$ has the smallest scaled stability threshold on the line, i.e.,

$$\frac{\lambda_m^*}{\nu_m} = \min_{i \in \{1,2,\ldots\}} \left[\frac{\lambda_i^*}{\nu_i}\right], \quad (21)$$

then $\mathrm{E}[T_i] = \nu_i \lambda \beta$ for all $i \neq m$ and for $\lambda < \lambda_m^*/\nu_m$. Therefore, given (20), $\lambda_m^*/\nu_m$ is the $\lambda$ that solves

$$\nu_m \lambda \beta = \gamma - \mathrm{E}[R_0] \prod_{i=1}^{m}(1-\sigma_i) - \sum_{j=1}^{m-1} \nu_j \lambda \beta \prod_{i=j+1}^{m}(1-\sigma_i). \quad (22)$$

Manipulating the equation above we obtain a closed form solution for the scaled stability threshold of station $m$ by

$$\frac{\lambda_m^*}{\nu_m} = \frac{\gamma - \mathrm{E}[R_0] \prod_{i=1}^m (1-\sigma_i)}{\sum_{j=1}^m \nu_j \beta \prod_{i=j+1}^m (1-\sigma_i)}, \quad (23)$$

where the product at the denominator is unitary for $j = m$. The solution does not depend on any capacity. Thus, it is not affected by the type of access control. To increase $\lambda_m^*/\nu_m$, we have to remove the condition (21), i.e., have at least a station $k$ with $\lambda_k^*/\nu_k < \lambda_m^*/\nu_m$. This proof only assumes the system to be stable and in stationary condition, and uses the relationship between the expected number of passengers entering cabins and the expected cabin capacity which has general validity. Passenger arrivals are not required to be Poisson distributed.

## VI. Ski Resort Performance Analysis

Let us analyze a stable system with at least one station in high load. If all stations were in low load, the system would be underutilized, thus passenger access control would not be needed. We reconsider part of the Bad Gastein resort with parameters given in Fig. 2. The equations describing the number of passengers entering cabins are:

$$\mathrm{E}[T_1] = \min[\nu_1 \lambda \beta, \eta_1] \quad (24)$$
$$\mathrm{E}[T_2] = \min[\nu_2 \lambda \beta, \gamma - \mathrm{E}[T_1](1-\sigma_2)] \quad (25)$$
$$\mathrm{E}[T_3] = \min[\nu_3 \lambda \beta, \gamma - \mathrm{E}[T_1](1-\sigma_2)(1-\sigma_3) \quad (26)$$
$$\qquad - \mathrm{E}[T_2](1-\sigma_3)].$$

Fig. 3a shows the impact of access control in Station 1 on the expected waiting time and the stability of Stations 2 and 3. The scaled stability thresholds $\lambda_m^*/\nu_m$ are shown as vertical lines. Quantities at the arrival rate of interest (near the instability of Station 2) are indicated by dots. The results can be interpreted as follows: Without access control at Station 1 ($\eta_1 = 8$), the expected waiting time $\mathrm{E}[W_2]$ at Station 2 is significantly worse than $\mathrm{E}[W_1]$ (see dots in Fig. 3a). Limiting the access at Station 1 to $\eta_1 = 7$ or $6$ — i.e. reserving one or two seats for Station 2 — marginally improves $\mathrm{E}[W_2]$ but significantly degrades the stability of Station 1 (the blue curve shifts to the left). Only limiting the access to $\eta_1 = 5$ creates a significant improvement of $\mathrm{E}[W_2]$ (the red curve shifts to the right). However, at this point, Station 1 is no longer stable at the arrival rate of interest.

Fig. 3b shows the expected number of passengers entering a cabin (accesses per service) over the arrival rate. To change the stability region of Station 2, we must shift the first non-differentiable point of $\mathrm{E}[T_1]$ to a value smaller than the first non-differentiable point of $\mathrm{E}[T_2]$, i.e., to make the stability region of Station 1 smaller than that of Station 2.

The same conclusion can be drawn from Fig. 4 (top plots): Decreasing the capacity of Station 1 by one or two seats ($\eta_1 = 7$ or $6$) does not change the scaled stability threshold $\lambda_2^*/\nu_2$, since the expected capacity $\mathrm{E}[C_2]$ does not change for $\lambda < \lambda_2^*/\nu_2$. Although some seats are reserved at Station 1, these are typically not used due to the low arrival rate. However, the limitation of $\mathrm{E}[C_1]$ leads to a gain in $\mathrm{E}[W_2]$ because of the reduction of variance $\mathrm{Var}[C_2]$ (third plot). We define the relative gain in waiting time as $h_m(i) = \frac{\bar{w}_m(8) - \bar{w}_m(i)}{\bar{w}_m(8)}$ with $\bar{w}_m(i) = \mathrm{E}[W_m]_{\eta_1 = i}$. In our scenario, for $\lambda_2$ near $\lambda_2^*$ and $i = 6$, this gain is about 50% (bottom plot): For example, if Station 2 is in high load and passengers wait about one hour, access control at Station 1 can reduce the average waiting time to half an hour. If Station 2 becomes unstable, the performance of Station 3 becomes a bit better. This is because the capacity distribution at Station 3 changes due to the instability of Station 2. Note that access control does not influence $\mathrm{Var}[C_3]$ and $\mathrm{E}[W_3]$ in Fig. 4. This is due to the high leaving probability $\sigma_3$. Stations are decoupled if $\sigma_3 = 1$.

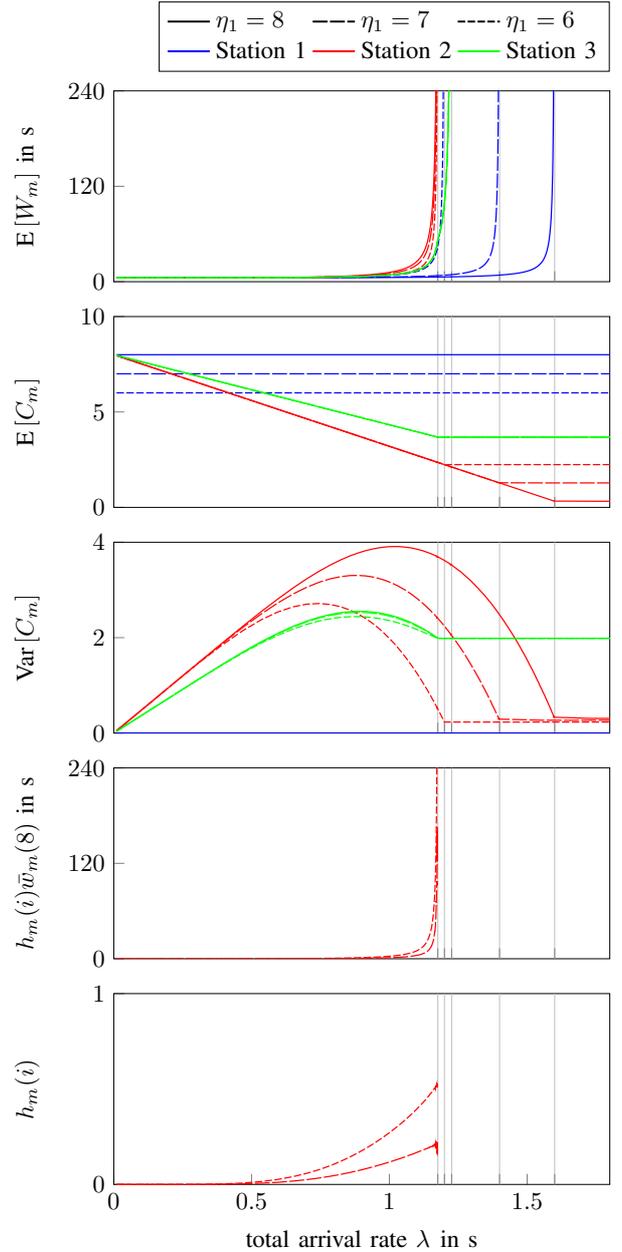

Figure 4: Capacity variance and gain in waiting time at Station 2. Parameters (same as Fig. 3): $\beta = 10$, $\gamma = 8$, $R_0 = 0$, $[\nu_1, \nu_2, \nu_3, \nu_4] = [0.5, 0.2, 0.3, 0]$, $[\sigma_1, \sigma_2, \sigma_3, \sigma_4] = [0, 0.04, 0.46, 1]$.

Fig. 5 shows the effect of capacity variance on the waiting time ($\sigma_m = 0$ and $\nu_m = 1/3$ $\forall m$). The capacity is varied by changing the distribution of $R_0$: $R_0^*(z)$ is $z^4$, $(z^2+z^6)/2$, $(z+z^7)/2$, $(1+z^8)/2$, with variances 0, 4, 9, and 16, respectively. We see that the variance of the capacity has an effect on the expected waiting time; this effect is transferred to other stations but it decreases with the number of stations. Therefore, to reduce the waiting time, good access control policies have to reduce the capacity variance.

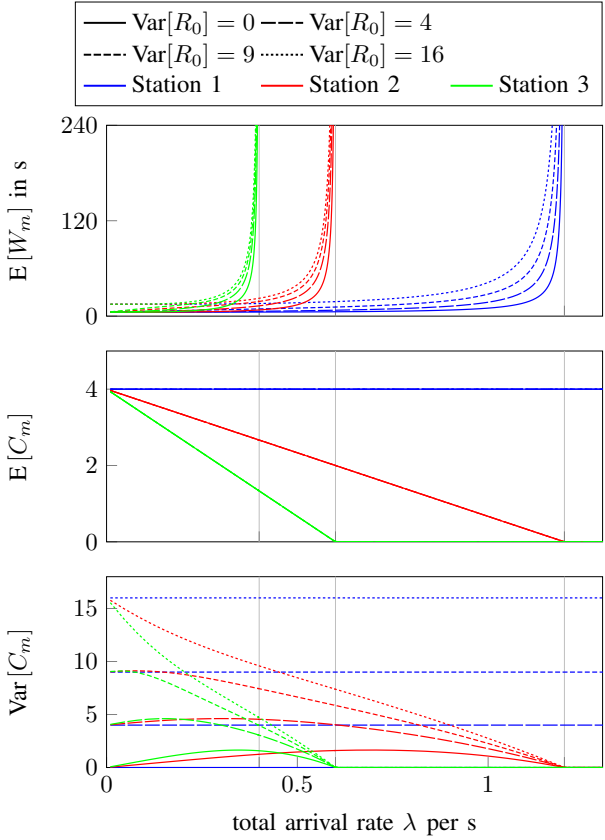

Figure 5: Effect of capacity variance on the expected waiting time. The capacity variance is modulated by changing the distribution of free seats at the ground station. $\beta = 10$, $\gamma = 8$, $E[R_0] = 4$, $R_0^*(z) = z^4, (z^2+z^6)/2, (z+z^7)/2, (1+z^8)/2$, $[\nu_1, \nu_2, \nu_3, \nu_4] = [1/3, 1/3, 1/3, 0]$, $[\sigma_1, \sigma_2, \sigma_3, \sigma_4] = [0, 0, 0, 1]$.

## VII. Conclusions: Toward Good Access Control

The main idea of access control is to limit the number of passengers entering certain stations so to increase the capacity of some targeted (succeeding) stations. In principle, an increase in the *average capacity* of a targeted station would increase the stability of that station and, in turn, decrease the waiting time by orders of magnitude. However, we proof that, in a linear topology and for a general arrival distribution, the average capacity of a station cannot be decreased without making one of the preceding stations less stable than the targeted station. Therefore, decreasing the average capacity is not a viable option. In addition, we show for Poisson arrivals that access control can decrease waiting time to an extend that is interesting for real operations, e.g., 50%. This decrease is due to a reduction of *capacity variance* at the targeted station. Even though the capacity variance is not a direct indicator of passengers' satisfaction, it plays an important role in the waiting time that passengers experience. Therefore, policies that aim to decrease the waiting time shall include the capacity variance in their performance indicators.

Our analysis of real passenger data in Bad Gastein shows that the arrival process is not a Poisson process. We conjecture that a greater variance in the arrivals (than in the Poisson case) directly effects the arrival station and indirectly the succeeding stations through the cabin capacity. In this case, a reduction of capacity variance becomes even more important. An algorithm based on this modeling, called *Gamora*, was tested on real passenger data from Bad Gastein. The algorithm uses the results on the scaled stability thresholds and on the expected cabin capacities, which are valid for a general distribution of passengers arrival. It shows very good performance and robustness with respect to estimation of passengers arrival and debording rate [16]. Experimental studies will follow.